\documentclass[11pt]{article}
\usepackage{moriond,epsfig}
\usepackage{floatflt}

\def\Journal#1#2#3#4{{#1} {\bf #2}, #3 (#4)}

\def\NPB{{\em Nucl. Phys.} B}
\def\PLB{{\em Phys. Lett.}  B}

\def\be{\begin{equation}}
\def\ee{\end{equation}}
\def\bea{\begin{eqnarray}}
\def\eea{\end{eqnarray}}

\begin{document}
\vspace*{4cm}
\title{(FINAL) HIGGS RESULTS FROM LEP}

\author{ U. SCHWICKERATH }

\address{CERN EP/DEE\\
1211 Geneva 23}

\maketitle\abstracts{
After the final end of data taking at LEP, the four collaborations continue
to analyze the large amount of data collected until Nov. 2000. The experiments
have published preliminary~\cite{ref:prelana} and final~\cite{ref:l3final} results
of the searches for the Standard Model (SM) Higgs boson within a few months after the
LEP shutdown. The combination of these results prefers the presence of a signal with a mass of about 
115.6~GeV/c$^2$, mainly due to an excess seen in the ALEPH data in that region.
More individual final results are available now, as well as a number of updates and
searches in different channels and models. This article summarizes the status of this field at the
time of the Moriond Conference 2002.
}

\section{Introduction}
\begin{floatingfigure}[r]{0.34\textwidth}
%\rule{5cm}{0.2mm}\hfill\rule{5cm}{0.2mm}
%\vskip 2.5cm
%\rule{5cm}{0.2mm}\hfill\rule{5cm}{0.2mm}
\psfig{figure=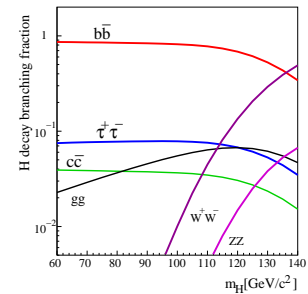,height=1.5in}
\caption{SM Higgs boson branching fractions.
\label{fig:brsm}}
\end{floatingfigure}

Despite of the big successes of the Standard Model, one of it's major
ingredients, the Higgs boson, has not yet been experimentally observered.
Electroweak precision data seems to prefer a light Higgs boson mass within
the reach of LEP~\cite{ref:blueband}.
Light Higgs bosons are also expected within a number of 
extensions of the SM. Within the MSSM, the light scalar Higgs boson is
expected to be lighter than about 150~GeV/c$^2$ \cite{ref:heinemeyer}.
The LEP collider offers a clean and powerful environment to look for such Higgs bosons.
Between the LEP startup in 1989 and it's final shutdown in November 2000 about 1000~$pb^{-1}$ of data were
delivered to each of the four experiments. Almost 700~$pb^{-1}$ per experiment were taken above
the $WW$ threshold. The analysis of this data is still going on, and
first final results on Higgs boson searches have been published. This presentation tries to summaries the status
of the search for Higgs bosons in a variety of different channels and models at the time of the
conference. All limits are given at 95\% CL.
\vspace*{0.4cm}

\section{The SM Higgs boson}
At LEP, the Higgs boson is expected to be produced mainly via the
Bjorken Strahlungs process. For final states with
two neutrinos or electrons, an additional small contribution to the production cross section
by $W$ or $Z$ fusion has to be taken into account.
At energies above the $Z$ threshold, final states are charac\-terized
by the decay products of an on-shell $Z$ boson, associated with two $b$ jets coming
from the Higgs decay. The expected branching fractions of a SM Higgs boson, calculated with HZHA~\cite{ref:hzha},
are shown in figure~\ref{fig:brsm}. 
As a consequence of the $Z$ branching fractions, the
most important search channels are the fully hadronic ($b\bar bq\bar q$) and
the missing energy channel ($b\bar b\nu\bar \nu$). Although final states with
charged leptonic decays ($b\bar bl^+l^-$) of the $Z$ are less frequent, they offer very
nice and clean final states for the search for the Higgs boson.

\subsection{Experimental results}
The SM Higgs analyses are optimized for Higgs bosons decaying into $b$ quarks, and thus use $b$ identification
as a major tool to suppress background from signal. The final states considered
can be characterized as $Hq\bar q$,$H\nu\nu$,$He^+e^-$, $H\mu^+\mu^-$ and $H\tau^+\tau^-$.
Signal efficiencies are estimated from simulation. $H\rightarrow \tau^+\tau^-$ decays are treated by a different 
ana\-lysis stream, looking for $\tau\tau Z$ final states, to increase the sensitivity of the search.

%\newpage
\subsection{SM combined results}
\begin{floatingfigure}[r]{0.35\textwidth}
%\rule{5cm}{0.2mm}\hfill\rule{5cm}{0.2mm}
%\vskip 2.5cm
%\rule{5cm}{0.2mm}\hfill\rule{5cm}{0.2mm}
\psfig{figure=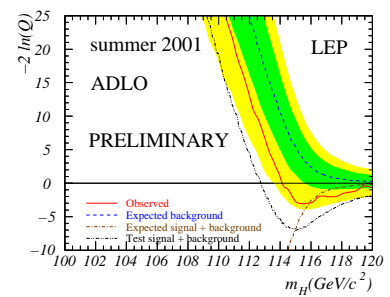,height=1.5in}
\caption{$-2lnQ$ distribution for all 4 LEP experiments (right). At the time of this combination, only
L3 data were finalized.
\label{fig:smcomb}}
\end{floatingfigure}
At the time of the conference, only the results of ALEPH~\cite{ref:alephfinal} and
L3~\cite{ref:l3final} were final. The latest combination was done in summer 2001~\cite{ref:lepsm}, based on
preliminary results of ALEPH, DELPHI and OPAL~\cite{ref:prelana}, and final results from L3~\cite{ref:l3final}.
A description of the combination method can be found in~\cite{ref:alex}.
Fig.~\ref{fig:smcomb} shows the $-2ln(Q)$ distribution from this combination.
It shows a deviation from the SM background expectation at masses around 115~GeV of the order of $2.1\sigma$.
Most of this effect comes from the ALEPH data. Due to this excess, the preliminary combined observed
limit of 114.1~GeV/c$^2$ is somewhat below the expectation of 115.4/c$^2$. 

Final results of ALEPH were published in December 2001.
Two independent analysis streams were applied to the same data and confirm the excess.
Assuming the absence of a signal,
ALEPH alone sets a limit of 111.5~GeV/c$^2$ on the mass of the SM Higgs boson, while the
expected limit is at 114.2~GeV. The minimal background probability, $1-CL_b$, of 2.4$\cdot10^{-3}$
is reached for $m_H\approx 115$~GeV/$c^2$~\cite{ref:alephfinal}. The change of these numbers with respect to
the preliminary result is of the order of a few hundred MeV/c$^2$. 

\vspace*{0.4cm}

\subsection{Flavor independent Higgs searches}
The standard technique to look for the SM Higgs boson makes use of presence of $b$ quarks in
the final states. The $b$ identification
turns out to be a powerful tool to suppress background in particular for Higgs masses close to $M_Z$.
However, in some models beyond the SM the decay of the Higgs into $b$ quarks can
be suppressed, which degrades the sensitivity of the standard search strategies.
Flavor independent searches drop the requirement of large $b$ significance.
As a consequence, the background in the 4-jet channel increases, and the missing energy channel
increases it's significance as compared to the SM analyses.  
No indication of Higgs boson production is observed, and a preliminary combined limit is set at 112.9~GeV/$c^2$
while 113.0~GeV/c$^2$ was expected~\cite{ref:lepflav}. One should note that this limit is
remarkably close to the SM result quoted above. It is valid for Higgs decays into gluons and all kind of quarks.

OPAL presented a preliminary analysis which goes one step further, and is completely independent of
the decay mode of the scalar boson. It combines all available LEP data, and excludes any kind of scalar boson
which is produced by the Bjorken Strahlungs process, between masses of 1keV/c$^2$ and 81~GeV/c$^2$~\cite{ref:opalsboson}. 

\subsection{Photonic Higgs decays}
Already within the SM a small fraction of Higgs bosons are predicted to decay
via a triangular loop into two photons. This topology requires a special treatment.
Preliminary analyses of DELPHI, L3 and OPAL explicitly look for final states $Z\rightarrow
l^+l^-,\nu\bar \nu,q\bar q$ with additional two photons. The ALEPH analysis focuses on the presence of
two energetic photons in the final state. A benchmark fermio-phobic model is used to derive mass limits in  this topology,
which assumes SM Higgs decays, but sets all couplings of the Higgs to fermions to zero.
That way, a mass limit of 108.2~GeV/c$^2$ can be derived~\cite{ref:lepphot}. The expected limit is 109~GeV/c$^2$.

\section{Invisible Higgs decays}
Some models suggest that the Higgs boson decays into particles which escape detection. An example
are Higgs decays into neutralinos, which can occur in some regions of the MSSM parameter
space, and can even become dominant. The signature of this channel are events with large missing energy and momentum,
associated with the decay products of a $Z$ boson.
Assuming a branching ratio of 100\% into invisible
particles, a preliminary combined limit of 114.4~GeV/c$^2$ can be set, to be compared with
113.6~GeV/c$^2$ expected~\cite{ref:lepinv}.

The ALEPH collaboration has finalized their results in this
channel~\cite{ref:alephfinal}.
Despite of small changes in the analysis, the numerical result remains unchanged with respect to the preliminary
analysis which was used in the last combination.

\section{MSSM Higgs bosons}
\subsection{Neutral Higgs bosons}
In the Minimal Supersymmetric Model (MSSM) additional Higgs bosons are predicted. Two scalar
Higgs bosons, $h$ and $H$, are produced via Strahlung or fusion, but with production
rates suppressed by $\sin(\alpha-\beta)$ and $\cos(\alpha-\beta)$, respectively, as compared to the SM Higgs boson. Final states
are very similar to the SM final states, and the same analyses can be applied.

\begin{floatingfigure}[r]{0.65\textwidth}
%\rule{5cm}{0.2mm}\hfill\rule{5cm}{0.2mm}
%\vskip 2.5cm
%\rule{5cm}{0.2mm}\hfill\rule{5cm}{0.2mm}
\psfig{figure=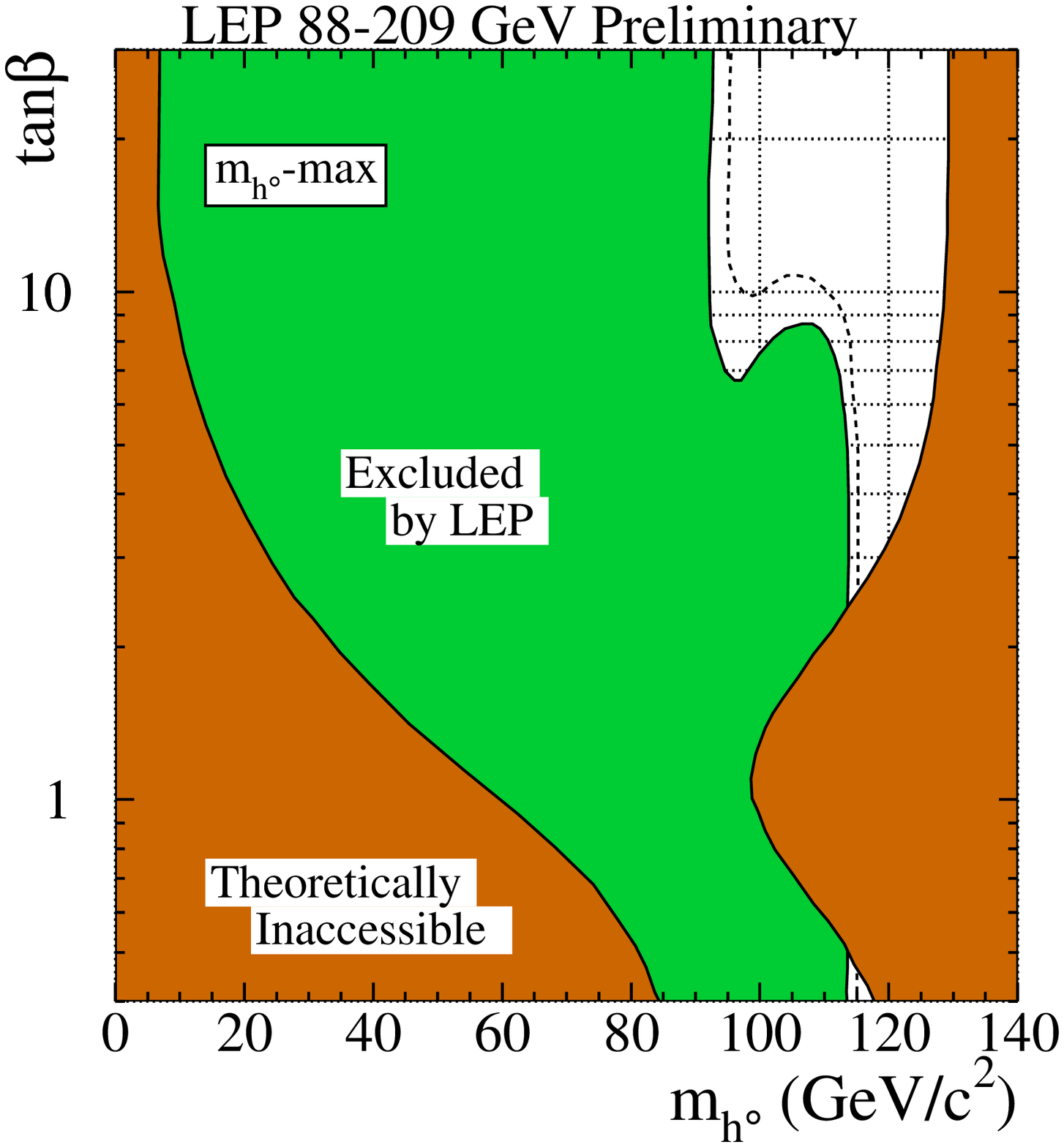,height=1.7in}\hfill
\psfig{figure=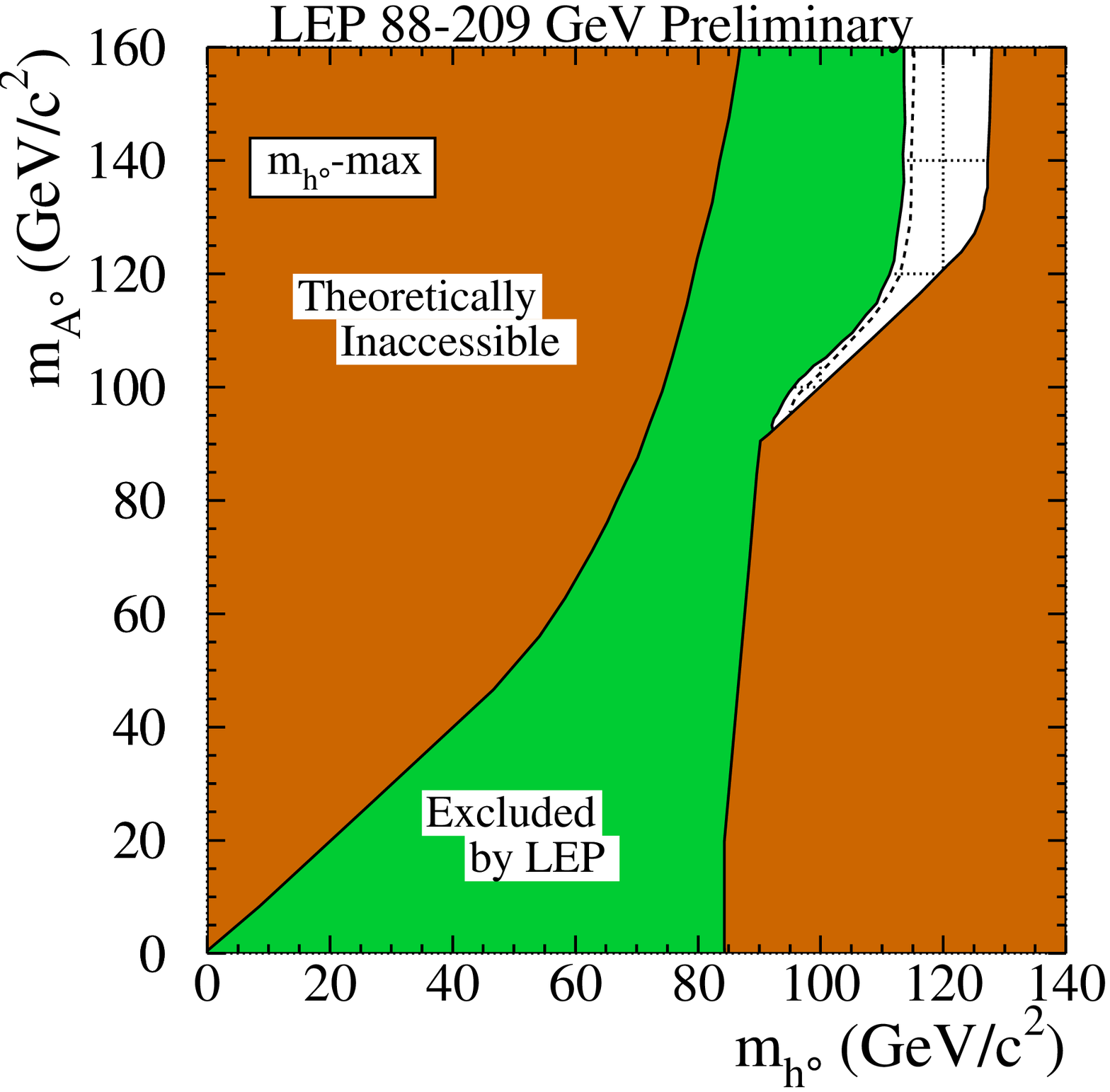,height=1.7in}
\caption{Combination of preliminary search results for neutral MSSM Higgs bosons, using the $m_h^{max}$ scenario.
\label{fig:mssmcomb}}
\end{floatingfigure}

The pseudo scalar Higgs boson, $A$, is expected to be produced in association with one of the
light scalar Higgs bosons. In typical mo\-dels, both Higgs bosons decay to a large fraction
into $b$ quarks, resulting in $b\bar bb\bar b$ events with a large charged mul\-ti\-pli\-ci\-ty, four jets
and large $b$ content. Less important but still significant are Higgs decays into $\tau$ leptons.
The resulting $b\bar b\tau^+\tau^-$ topologies require dedicated analyses.

At the time of the conference, all experiments had preliminary results for neutral MSSM
Higgs bosons. The combined limits from these results, interpreted in
the $m_h^{max}$ scenario~\cite{ref:mhmax}, are shown in fig.~\ref{fig:mssmcomb} in the
$m_h$ - $\tan\beta$ and in the  $m_h$ - $m_A$ plane~\cite{ref:lepmssm}. The current limits on $m_h$ and
$m_A$ are at 91.0~GeV/c$^2$ and 91.9~GeV/c$^2$, respectively. From fig.~\ref{fig:mssmcomb}, the
region $0.5<\tan\beta<2.4$ is also excluded.

ALEPH final results on neutral MSSM Higgs bosons are published
together with the SM Higgs results~\cite{ref:alephfinal}. The difference with respect to the preliminary results is small,
and of the order of a few hundred MeV/c$^2$.

\subsection{Charged Higgs bosons}
Charged Higgs bosons in the MSSM are expected to be produced in pairs. They decay preferably like
$H^+\rightarrow c\bar s$ or $\tau^+\nu$, so that three possible final states have to be
considered. No indication for a signal has been seen, and the preliminary combined limit is
78.6~GeV/$c^2$ with 78.8~GeV/$c^2$ expected~\cite{ref:lephphm}. 

\section{Doubly charged Higgs bosons}
Doubly charged Higgs bosons are predicted by some left-right symmetric models. They are
expected to be produced in pairs via $s$-channel $Z/\gamma$ exchange, or $t$-channel charged
lepton exchange. Due to charge conservation, they are expected to decay into leptons,
preferably into $\tau$ leptons. 
New results were presented on the search for doubly charged Higgs boson by DELPHI and OPAL.
The preliminary DELPHI analysis~\cite{ref:delphihpphmm} concentrates on 4$\tau$ final states, and sets a limit of 99.1~GeV/$c^2$.
The final OPAL analysis~\cite{ref:opalhpphmm} sets a limit at 98.5~GeV/$c^2$. The latter analysis represents the  first search in
this channel at energies above $M_Z$, and includes all possible final states.

\section*{Acknowledgments}
I want to thank the four LEP experiments and the LEP Higgs working group for providing information
to prepare this presentation.

\end{document}